# Deep learning segmentation of fibrous cap in intravascular optical coherence tomography images

**Juhwan Lee[1], Justin N. Kim[1], Luis A. P. Dallan[2], Vladislav N. Zimin[3], Ammar Hoori[1], Neda S. Hassani[2], Mohamed H. E. Makhlouf[2], Giulio Guagliumi[4], Hiram G. Bezerra[5], David L. Wilson[1,6,*]**

[1] Department of Biomedical Engineering, Case Western Reserve University, Cleveland, OH, 44106, USA
[2] Harrington Heart and Vascular Institute, University Hospitals Cleveland Medical Center, Cleveland, OH, 44106, USA
[3] Brookdale University Hospital Medical Center, 1 Brookdale Plaza, Brooklyn, NY, 11212, USA
[4] Cardiovascular Department, Galeazzi San'Ambrogio Hospital, Innovation District, Milan, Italy
[5] Interventional Cardiology Center, Heart and Vascular Institute, University of South Florida, Tampa, FL, 33606, USA
[6] Case Western Reserve University, Department of Radiology, Cleveland, OH, 44106, USA

*Corresponding author: dlw@case.edu
Telephone number: 216-368-4099, fax: 216-368-4969

## Abstract

**Background and Objective:** Thin-cap fibroatheroma (TCFA) is a prominent risk factor for plaque rupture. Intravascular optical coherence tomography (IVOCT) enables identification of fibrous cap (FC), measurement of FC thicknesses, and assessment of plaque vulnerability. We developed a fully-automated deep learning method for FC segmentation.

**Methods:** This study included 32,531 images across 227 pullbacks from two registries (TRANSFORM-OCT and UHCMC). Images were semi-automatically labeled using our OCTOPUS with expert editing using established guidelines. We employed preprocessing including guidewire shadow detection, lumen segmentation, pixel-shifting, and Gaussian filtering on raw IVOCT $(r, \theta)$ images. Data were augmented in a natural way by changing $\theta$ in spiral acquisitions and by changing intensity and noise values. We used a modified SegResNet and comparison networks to segment FCs. We employed transfer learning from our existing much larger, fully-labeled calcification IVOCT dataset to reduce deep-learning training. Postprocessing with a morphological operation enhanced segmentation performance.

**Results:** Overall, our method consistently delivered better FC segmentation results (Dice: 0.837±0.012) than other deep-learning methods. Transfer learning reduced training time by 84% and reduced the need for more training samples. Our method showed a high level of generalizability, evidenced by highly-consistent segmentations across five-fold cross-validation (sensitivity: 85.0±0.3%, Dice: 0.846±0.011) and the held-out test (sensitivity: 84.9%, Dice: 0.816) sets. In addition, we found excellent agreement of FC thickness with ground truth (2.95±20.73 $\mu m$), giving clinically insignificant bias. There was excellent reproducibility in pre- and post-stenting pullbacks (average FC angle: 200.9±128.0° / 202.0±121.1°).

**Conclusions:** Our fully automated, deep-learning FC segmentation method demonstrated excellent performance, generalizability, and reproducibility on multi-center datasets. It will be useful for multiple research purposes and potentially for planning stent deployments that avoid placing a stent edge over an FC.

# 1    Introduction

Thin-cap fibroatheroma (TCFA) is widely recognized as a prominent risk factor for plaque rupture, a major contributor to acute coronary syndromes (ACS) [1,2]. TCFA is typically characterized by the presence of a large lipid pool covered by a thin fibrous cap (FC) (<65 $\mu m$) and increased macrophage activity [1,3]. However, Kume et al. reported thicker cap measurements using intravascular optical coherence tomography (IVOCT), possibly due to tissue shrinkage during histology [4]. The consensus standard also recommends adjusting this threshold when applied to IVOCT images to account for potential tissue shrinkage (10-20%) during histopathologic processing [5]. Intravascular imaging techniques such as intravascular ultrasound (IVUS) and IVOCT enables the assessment of FC thickness. Nevertheless, caution should be exercised when interpreting IVUS findings due to insufficient resolution (150-200 $\mu m$) for reliable thickness measurement. In contrast, IVOCT offers superior axial resolution (12-18 $\mu m$) [6], facilitating precise measurement of FC thickness, identification of TCFA, and determination of plaque vulnerability [5].

Despite its advantages in assessing thin FC tissues, IVOCT presents significant limitations for real-time treatment planning and research in large data sets. First, an IVOCT pullback typically consists of more than 300 image frames, resulting in a data overload. The comprehensive manual analysis of coronary plaques necessitates meticulous consideration of image characteristics, leading to a time-consuming and labor-intensive process. Second, manual analysis of IVOCT images can be prone to high levels of inter- and intra-observer variability [5]. For instance, our research group reported intra- and inter-observer variabilities of ≤5% and 6%, respectively, among experienced cardiologists in detecting stent struts in IVOCT images [7]. Since coronary plaques exhibit less distinct features compared to stent struts, the variability in plaque analysis among clinicians is expected to be even higher. An automated method will be more reproducible which will be especially important if one is looking for changes between cohorts or within a cohort as with a drug trial. Third, manual point measurements of fibrous cap thickness do not fully capture tendency to rupture. Automated assessment of the surface area of a lesion is surely needed to assess vulnerability. These observations underscore the imperative for an accurate and fully-automated method for fibrous cap analysis.

There have been only a few studies addressing these limitations. Our group initially developed a semi-automated method for volumetric quantification of FC [8]. Briefly, we segmented the luminal and abluminal boundaries in the polar coordinates of IVOCT images using dynamic programming. Then, we quantified the thickness at each point of the FC luminal boundary. Although the method was validated in various ways, the manual identification of the circumferential distribution of the lipid was required. Zahnd et al. [9] proposed a semi-automatic segmentation method utilizing dynamic programming to quantify coronary FC thickness in IVOCT images. The method was evaluated through multiple approaches, and the results were promising. However, its major limitation was the requirement of manual initialization of lipidic plaque for fibrous cap segmentation. Additionally, the method was validated on a small subset of pullbacks (179 images from 21 patients), which reduces its reproducibility. Min et al. [10] developed a deep learning model using a DenseNet architecture for classifying IVOCT frames as either TCFA or non-TCFA. They reported a promising classification performance with an overall accuracy of 91.6±1.7%, sensitivity of 88.7±3.4%, and specificity of 91.8±2.0%. However, this method did not offer quantitative measurements of FC. In our previous study [11], we developed an automated method capable of detecting lipidic plaque and segmenting the FC in IVOCT images. The method consisted of two phases: lipidic plaque detection using deep learning and FC segmentation using dynamic programming. Evaluation was performed on over 4,000 image frames from 41 patients, demonstrating excellent discriminability of lipidic plaque and good reproducibility in FC thickness measurement. However, our previous method occasionally exhibited inaccurate lipid arc detections, affecting the accuracy of FC thickness measurement. An accurate and fully automated end-to-end training method could offer faster and improved assessment of FC.

In this report, we expand upon our previous study [11] and develop a fully-automated deep learning method for FC segmentation in IVOCT images. To achieve this, we employ specialized image preprocessing, transfer learning, a large carefully labeled dataset, physics- and system-plausible augmentation, and advanced deep learning networks. Robustness, accuracy, and reproducibility of results are carefully evaluated. Because we are automatically analyzing volumes of data, we create visual heatmaps of fibrous cap thickness, giving a compelling visualization of vulnerability.

## 2    Image Analysis Methods

### 2.1 Preprocessing

We employed a preprocessing method previously proposed by our group [12,13] to identify the appropriate tissue regions of interest for FC segmentation. Preprocessing of raw IVOCT ($r,\theta$) image data involved several steps: (1) Detection and removal of the guidewire and corresponding shadow regions using dynamic programming [14], as they do not contain tissue information. (2) Segmentation of the lumen boundary using a deep learning-based semantic segmentation method developed by our group [15]. (3) Pixel-shifting each A-line to the left, ensuring that all A-lines have the same starting point along the radial direction. This step was crucial as it not only created a smaller region of interest for deep learning, simplifying the processing, but also aligned the tissues, making different lesions appear more similar to the network [11]. (4) Limiting the $r$ direction to the first 200 pixels (~1 $mm$) due to the limited penetration depth of the IVOCT signal. (5) Applying a Gaussian filter with a kernel size of (7,7) and a standard deviation of 1-pixel to reduce noise. After preprocessing, the size of the IVOCT data was reduced from (968×448 pixels) to (200×448 pixels) without any loss of meaningful data. The preprocessed images were then used for further processing. The overall workflow of the proposed method is illustrated in Figure 1. Please note that all images in the manuscript are presented following a log transformation for improved visualization.

### 2.2 Data augmentation for deep learning

Our previous study found that data augmentation significantly improved deep learning segmentation performance in IVOCT images [15,16]. The IVOCT data were augmented for deep learning training to increase the number of examples with varying FC locations and intensities, thereby enhancing the spatial invariance of the methods. We applied two data augmentation approaches, one for the raw polar IVOCT before preprocessing and another for the preprocessed IVOCT images. First, for the raw polar IVOCT pullback, we concatenated all of the raw polar ($r,\theta$) images to form one large 2D array, where $r$ represents tissue depth and the $\theta$ is catheter rotation, which rotates from 0 to $N \times 360°$, where $N$ is the number of unput images. Then, we changed the offset angle to extract new polar image frames with no data loss or distortion. In practice, we shifted the starting A-line six times by 80 A-line increments. Further details are provided elsewhere [15,16]. Second, the data augmentation for preprocessed IVOCT data involved several steps as follows: (1) We normalized the intensity of all input images to the range (0, 1). (2) We flipped input images along the vertical axis with a 10% probability. (3) We randomly scaled the pixel intensity of the input image by a factor of 0.1 with a 20% probability. (4) We randomly shifted the pixel intensity of the input image by a factor of 0.1 with a 20% probability.

### 2.3 FC segmentation

To segment FC regions in IVOCT images, we utilized a modified version of SegResNet [17], which follows an encoder-decoder-based convolutional neural network (CNN) architecture with an asymmetrically large encoder backbone and a smaller decoder (Fig. 2). In our study, the variational auto-encoder branch was excluded since we had a sufficient number of IVOCT image instances (>32,000) for FC segmentation.

The encoder component primarily consisted of ResNet [18] blocks, where each block comprised two convolutions with normalization and rectified linear unit activation, followed by an additive identity skip connection. For normalization, we employed Group Normalization [19], which demonstrates improved performance compared to Batch Normalization when the batch size is small. We adopted a conventional CNN approach to progressively downsize the image dimensions by a factor of 2 while simultaneously increasing the feature size by 2. Strided convolutions were employed for downsizing, and all convolutions were 3×3 with an initial number of filters set to 16. Additionally, a dropout layer with a probability of 0.2 was incorporated into each block.

The decoder component resembled the encoder part but contained only a single block for each spatial level. Each decoder level commenced with upsizing, which reduced the number of features by a factor of 2 using 1x1 convolutions and doubled the spatial dimension via bilinear upsampling. Subsequently, the encoder output from the corresponding spatial level was added. The output of the decoder retained the same size as the original image, and the number of features matched the initial input feature size. Finally, a 1x1 convolution layer and a sigmoid function were applied.

*2.4 Transfer learning for deep learning*

To optimize the deep learning training for FC segmentation, we employed transfer learning with our existing calcification IVOCT dataset. The rationale behind transfer learning is to reduce the training time and the need for a large number of training samples by utilizing a model that has been trained on a different but related task, specifically calcification segmentation in IVOCT images. For this study, we constructed a pretrained network with the same architecture as the FC segmentation model using the IVOCT data from our previous studies on calcification segmentation (comprising over 24,000 images) [13,15,20,21]. Rather than initializing the weights of the second network (FC segmentation) randomly, we utilized the pretrained model to initialize these weights. Through transfer learning, the model can rapidly reach the convergence point during training, potentially leading to enhanced performance.

*2.5 Postprocessing*

To clean results and enhance segmentation performance, we implemented a morphological operation after the FC segmentation. Given the presence of inherent speckle noises in IVOCT images, the network occasionally exhibits spotty segmentation errors throughout the pullbacks. We employed an opening operation on the output labels with a disk-shaped structuring element with a radius of 3. Subsequently, we filled in the holes within the segmented labels. The pixel connectivity rule was set to 4.

# 3    Experimental Methods

*3.1 Data acquisition*

The images utilized in this study were obtained from two sources: the TRiple Assessment of Neointima Stent FOrmation to Reabsorbable polyMer with Optical Coherence Tomography (TRANSFORM-OCT) trial [22] and the University Hospitals Cleveland Medical Center (UHCMC) Registry. The TRANSFORM-OCT dataset comprised 24,209 images (15,239 calcification and 8,970 FC images) derived from 153 pullbacks involving 77 patients. On the other hand, the UHCMC dataset consisted of 8,322 images (6,960 calcification and 1,362 FC images) acquired from 74 pullbacks involving 74 patients. The raw IVOCT data size was $968 \times 448$ in the $(r, \theta)$ domain. Calcification images were employed to establish the pretrained network for transfer learning, whereas FC images were employed for training the network specifically for FC segmentation (Fig. 1). The IVOCT images were acquired using a frequency-domain ILUMIEN OCT system (Abbott Vascular, Santa Clara, CA, USA), which utilized a tunable light source sweeping from 1,250 to 1,360 *nm*. The imaging pullback was conducted at a frame rate of 180 fps, pullback speed of 36 *mm/s*, and an axial resolution of approximately 20 *μm*.

The inclusion criteria encompassed patients with stable angina and documented ischemia or acute coronary syndrome who had undergone IVOCT examination. Major exclusion criteria included the presence of unprotected left main disease, chronic total occlusion, baseline serum creatinine >2.0 *mg/dL*, life expectancy <18 months, and unsuitability for OCT imaging at the clinician's discretion. This study adhered to the principles outlined in the Declaration of Helsinki and received approval from the Institutional Review Board of University Hospitals Cleveland Medical Center, Cleveland, OH, USA.

*3.2 Ground truth labeling*

The FC regions were segmented using Optical Coherence TOmography PlaqUe and Stent (OCTOPUS) software, previously developed by our group [23], and manually edited by two experts (5+ years of experience) from the Cardiovascular Imaging Core Laboratory at University Hospitals Cleveland Medical Center, a leading IVOCT analysis laboratory with over 3,000 clinical trial cases analyzed. For manual editing of FC, we followed the "consensus document" for IVOCT image analysis [24]. Specifically, the FC was defined as a distinct tissue layer of connective tissue, which is often signal-rich, overlying a signal-poor region, and TCFA was defined as a necrotic core with an overlying fibrous cap where the minimum thickness of the fibrous cap was less than a predetermined threshold (<65 *μm*). The labels provided by the more experienced expert were used as the ground truth. In case of disagreement between the two readers, they reevaluated the frames and reached a consensus decision. Any region other than FC was given a label of "background", which allowed us to set up a binary segmentation task.

*3.3 Network training*

We utilized the AdamW optimizer, an adaptive moment estimation optimizer with decoupled weight decay [25], for training both the transfer learning and FC segmentation networks. This optimizer employs stochastic gradient descent and adaptively estimates first-order and second-order moments while incorporating a weight decay method [25]. The AdamW optimizer is computationally efficient, robust to diagonal rescaling of gradients, and well-suited for handling large-scale data problems.

The initial learning rate, epsilon, and weight decay were empirically set to $1 \times 10^{-5}$, $1 \times 10^{-9}$, and $1 \times 10^{-6}$, respectively. To train the networks, we employed a maximum of 600 epochs and a batch size of 64. L2 norm regularization with a weight of $1 \times 10^{-6}$ was applied to the convolutional kernel parameters, and spatial dropout with a rate of 0.2 was implemented after the initial encoder convolution, following the original implementation [17].

For the FC segmentation model, the learning parameters of each encoder layer were initialized by transferring weights from the transfer learning network, which was pretrained on IVOCT calcification data. The network weights were then fine-tuned in a layer-by-layer manner using backpropagation. The learning rates of subsequent layers were adjusted sequentially until the performance on the validation set ceased to improve.

The loss functions for both networks were computed using the Dice loss function over the softmax outputs. To prevent overfitting during training, we employed a stopping criterion that halted training when the validation loss failed to improve for 10 consecutive epochs or when the maximum number of epochs was reached. In practice, the former rule was executed. We used the following frameworks using Python (ver. 3.9.13, Python Software Foundation, USA): Pytorch (ver. 1.13.1) and Monai (ver. 1.1.0).

*3.4 Performance evaluation*

For transfer learning training, we partitioned a total of 227 pullbacks into training and validation sets. Following a 7:3 split, the training set consisted of 15,239 calcification images from 153 pullbacks (TRANSFORM-OCT), while the validation set contained 6,960 calcification images from 74 pullbacks (UHCMC). There was no held-out test set for pretraining as the network was solely employed for transfer learning purposes.

Regarding FC segmentation, we performed a five-fold cross-validation on the TRANSFORM-OCT dataset, which encompassed 8,970 FC images from 153 pullbacks. In each fold, there were sub-groups for training (60%), validation (20%), and testing (20%). Folds and sub-groups were based on pullbacks rather than images. This approach ensured that each sub-group was assigned to the test set precisely once, thereby avoiding evaluation variance. Additionally, the UHCMC dataset, comprising 1,362 FC images from 74 pullbacks, served as the held-out test set for further evaluations.

The segmentation performance was quantitatively assessed using conventional metrics, including pixel-wise positive predictive value (PPV), negative predictive value (NPV), sensitivity, specificity, and Dice coefficient as below:

$$PPV = TP / (TP + FP) \tag{1}$$

$$NPV = TN / (TN + FN) \tag{2}$$

$$Sensitivity = TP / (TP + FN) \tag{3}$$

$$Specificity = TN / (TN + FP) \tag{4}$$

$$Dice\ coefficient = 2TP / (2TP + FP + FN) \tag{5}$$

Here, TP represents the number of true positive pixels, TN denotes the number of true negatives, FP signifies the number of false positives, and FN represents the number of false negatives. We reported the mean and standard deviation of these metrics across the five folds. Furthermore, to investigate performance variance, we compared the segmentation results against three other networks (U-Net [26], Attention U-Net [27], and nnU-Net [28]). We also examined the impact of transfer learning on the results by comparing performance with and without transfer learning. To conduct this analysis, we created pretrained networks for all four networks, utilizing the exact same calcification-analysis training and validation datasets. Additionally, we assessed the reproducibility of the proposed method by evaluating pre- and post-stenting IVOCT pullbacks acquired from the same lesion.

In addition to conventional metrics, we also assessed clinically meaningful metrics, namely FC thickness, which has been previously utilized in several clinical research studies [29,30]. This metric was measured on the held-out test set. FC thickness refers to the distance between the luminal boundary and the abluminal boundary. The distance was measured at each point along the abluminal boundary from the luminal boundary in polar $(r,\theta)$ images.

## 4    Results

For FC segmentation, SegResNet consistently delivered the most reliable segmentation results as compared to other deep learning networks (Fig. 3 and Table 1). SegResNet segmentations are nearly visually identical to manual labels (Fig. 3). Attention U-Net exhibited the lowest Dice coefficient (0.806±0.022) and sensitivity (80.1%±4.9%) among all the networks (Table 1). The nnU-Net showed a slightly higher (or equivalent) sensitivity (91.4%±1.5%) and Dice coefficient (0.837±0.008) than the SegResNet. However, its PPV was the lowest (77.2%±0.3%), indicating a higher rate of false positives. Overall, the SegResNet demonstrated the best segmentation performance out of all the networks employed. All networks underwent the same preprocessing, data augmentation, transfer learning, and postprocessing. As described previously, networks were pre-trained on the task of segmenting calcifications, using our large database of such images. For more information on pre-training, see Supplementary Fig. S1 and Table S1.

Transfer learning from the calcification segmentation task reduced training time without yielding significant improvements in FC segmentation performance. Figure 4 shows mean validation Dice loss curves with and without transfer learning from SegResNet. Without transfer learning, convergence (as determined by the stopping rule) required approximately 22-42 epochs, whereas with transfer learning, convergence was achieved in only 3-7 epochs. Transfer learning also reduced the need for a large number of training samples (Fig. 5). With transfer learning, Dice values always exceed results without transfer learning. They appear to reach asymptotic convergence unlike without transfer learning. With transfer learning, only 60% of samples gives a better result than 100% of the samples without transfer learning. Using the full data set and long training times, transfer learning resulted in somewhat improved quantitative metrics across the folds for all networks, as shown in Supplementary Table S2. If we reduce the training samples by 40%, then there is a significant difference between with and without transfer learning (not shown).

Our method had highly consistent segmentation results between the five-fold cross-validation (TRANSFORM-OCT) and the held-out test (UHCMC) sets (Fig. 6 and Table 2). To achieve this, we applied each of the five trained models from the cross-validation on the held-out test set, which consisted of 1,362 FC images from 74 pullbacks at UHCMC. We combined the pixel-wise predictions by selecting the most common output through plurality voting. On the held-out set, the proposed method achieved a PPV of 78.5%, sensitivity of 84.9%, and Dice coefficient of 0.816. These metrics indicate a high level of generalizability for our method.

We found very good agreement of mean FC thickness measurements between our automated method and manual ground truth assessments (Fig. 7). The linear regression analysis showed an $R^2$ of 0.909, indicating significant correlations between the ground truth and the proposed method (Fig. 7(A)). The mean bias of FC thickness measurements was only about 2.95±20.73 $\mu m$ in a Bland-Altman analysis, and most measurements (98%, 1330/1362) were included within the limits of agreement (Fig 7(B)). These results support no significant bias of the proposed method as compared to the ground truth.

Our method demonstrated excellent reproducibility of automated FC segmentation in scan-rescan IVOCT images (Fig. 8). For this analysis, we utilized IVOCT images from paired pre- and post-stenting IVOCT pullbacks outside the stented region. We extracted 51 paired IVOCT images containing FCs, which were not included in the training or held-out test sets. Lesion attributes between pre- and post-stenting pullbacks were: average FC thickness (87.6±38.6 $\mu m$ / 105.8±33.9 $\mu m$), average FC arc angle (200.9±128.0° / 202.0±121.1°), average FC area (1.04±0.62 $mm^2$ / 1.01±0.56 $mm^2$), and FC surface area (8.6 $mm^2$ / 7.5 $mm^2$). Additionally, the coefficients of variation between pre- and post-stenting pullbacks were very similar (Supplementary Table S3), indicating the strong reproducibility of our method.

Figure 9 depicts 3D visualizations of FC thickness on representative IVOCT pullbacks with large and small lesions (Fig. 9). For both lesions, the minimum thickness was under 65 $\mu m$, making both lesions TCFAs under the normal definition. However, from biomechanics, the large lesion will tend to undergo much larger strains than the smaller one due to the presence of surrounding normal support tissues on the smaller one. Presumably the larger lesion will be much more prone to rupture.

## 5    Discussion

In this study, we built on our previous studies of IVOCT image analysis [8,11–13,15,16,20,21,23,29,31–34] and developed a fully-automated method for FC segmentation in IVOCT images using deep learning. The main findings are: (1) Transfer learning from a separate calcification segmentation task greatly reduced both the training time and the need for a large number of labeled training samples for deep learning. (2) The SegResNet network outperformed other deep learning networks, including U-Net, Attention U-Net, and nnU-Net. (3) Our method had a high generalizability evidenced by similar segmentation results on training (TRANSFORM-OCT) and held-out test (UHCMC) sets. (4) Our method showed a good reproducibility for FC segmentation in scan-rescan IVOCT pullbacks. (5) The proposed method produced excellent results for FC segmentation while taking a reasonable amount of time (0.02s per frame) to compute, implying that it could be a promising solution for both research and clinical applications. (6) Using our method, it is possible to create 3D FC thickness heatmaps and histograms of FC thickness of FC lesions, creating a broader characterization of plaque vulnerability.

Transfer learning enabled much faster training as compared to the traditional methods without transfer learning across all the networks used in this study. Transfer learning leverages prior knowledge gained while solving one task to solve a related new task [35]. In this study, prior knowledge was obtained from previous calcification segmentation [13,15,20,21], and it was transferred to a related new task, which was FC segmentation. The whole training process can be made more efficient and generalizable by reusing elements of an algorithm and transferring the knowledge already held by a model. In addition, the sharing of knowledge between two different models can result in a more accurate and effective model overall. In our experiments, transfer learning achieved the desired performance in ~20% of the time. More importantly, with transfer learning, the network only required 60% of the training samples to reach the desired performance level. In our case, transfer learning was done from a similar task (calcification segmentation) to the target task (FC segmentation). We also tried pretraining using a conventional less related task (i.e., masked autoencoder). In our hands, using the more similar calcification task gave superior results.

Our method enabled excellent reproducibility of FC thickness measurement. From the experiments on the paired pre- and post-stenting IVOCT pullbacks, we found a very small bias and high similarity between those pullbacks (Fig. 8). IVOCT inevitably exhibits variations in plaque composition due to its reliance on the positioning of the catheter and guidewire, even when imaging the same lesion accurately. As a result, there were slight differences in the characteristics of FC plaque between the pre- and post-stenting pullbacks. Nonetheless, our automated analysis offered high repeatability, surpassing the potential variability introduced by different analysts, even with minimal user intervention. This suggests that our method is well-suited for large-scale research studies. Furthermore, with further improvements, it has the potential to be utilized clinically, assisting physicians in determining an appropriate stent landing zone.

We found an excellent generalizability of the proposed method for FC analysis using two large cohort data sets (i.e., TRANSFORM-OCT and UHCMC). Clinical research studies have mostly relied on single-center data, limiting their generalizability as they focus on local data despite the potential for larger datasets. Therefore, to generalize the clinical research methods (particularly deep learning applications), it is essential to use many data acquired from multiple sites. In our study, we created deep learning models using the TRANSFORM-OCT trial data, that mainly includes Italian patients, and validated its generalizability on the UHCMC data (Cleveland, OH, USA).

This study has identified some limitations. First, our method occasionally produced inaccurate results when there was a side branch or when there was mixed plaque (see Supplementary Fig. S2). However, such cases were infrequent and easily correctable. Second, although a large dataset was used in this study, further improvement may be attainable in future studies utilizing larger datasets. Some instances were probably not fully represented in training (e.g., cases with a side branch). Active learning or human-in-the-loop learning where cases are segmented, corrected, and then added back to the training mix, could possibly improve performance further. Third, as we have seen performance differences between networks, this suggests that more advanced deep learning approaches might further improve results.

In conclusion, we developed a fully automated, deep learning FC segmentation and measurement methods for IVOCT images. Using multi-center datasets, we performed rigorous evaluations and demonstrated excellent performance, generalizability, and reproducibility. We believe that this method will prove useful for various research applications and may even play a role in future treatment planning, especially when one wants to avoid placing a stent edge over a lipidic lesions with a vulnerable FC.

## Acknowledgments

This project was supported by the National Heart, Lung, and Blood Institute through grants NIH R21HL108263, NIH R01HL114406, and NIH R01HL143484. This research was conducted in space renovated using funds from an NIH construction grant (C06 RR12463) awarded to Case Western Reserve University. The content of this report is solely the responsibility of the authors and does not necessarily represent the official views of the National Institutes of Health. The grants were obtained via collaboration between Case Western Reserve University and University Hospitals of Cleveland. This work made use of the High-Performance Computing Resource in the Core Facility for

**Figures**

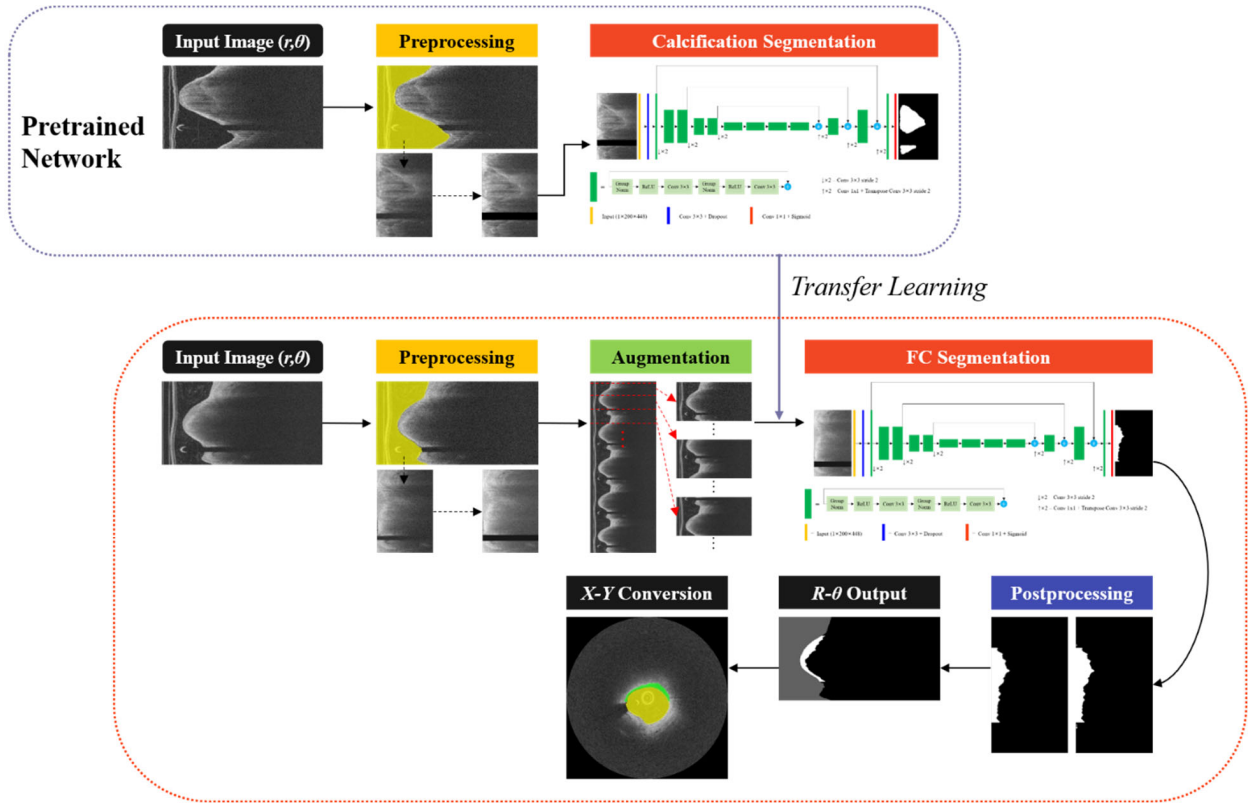

Fig 1. Workflow of FC segmentation in IVOCT images. The key steps include preprocessing, data augmentation, FC segmentation, transfer learning, and postprocessing. Preprocessing involves guidewire shadow detection, lumen segmentation, pixel-shifting, and noise filtering on raw IVOCT data (*r,θ*), followed by data augmentation on the preprocessed images. The output serves as input to the FC segmentation network. Postprocessing techniques, such as filling and morphological operations, are utilized to reduce small false positive errors. For transfer learning (top, blue dotted box), the network is trained using IVOCT calcification images with the same preprocessing and network architecture.

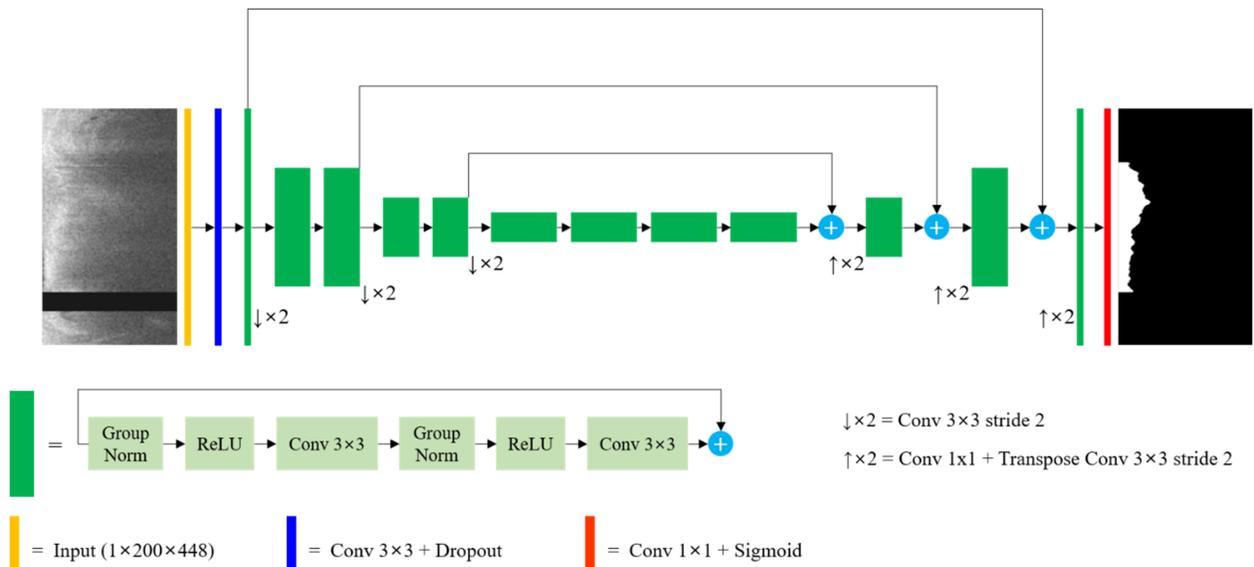

Fig 2. SegResNet Architecture for FC Segmentation. The preprocessed IVOCT image serves as the input, starting with an initial 3x3 convolution and dropout layers. Each green block represents a ResNet-like block with group normalization. The decoder outputs a predicted label, followed by a sigmoid activation function to generate a pixel-wise classification map. Both the input and output images have the same size (200x448 pixels in $(r,\theta)$). In the input image, the black strip indicates the removed guidewire shadow.

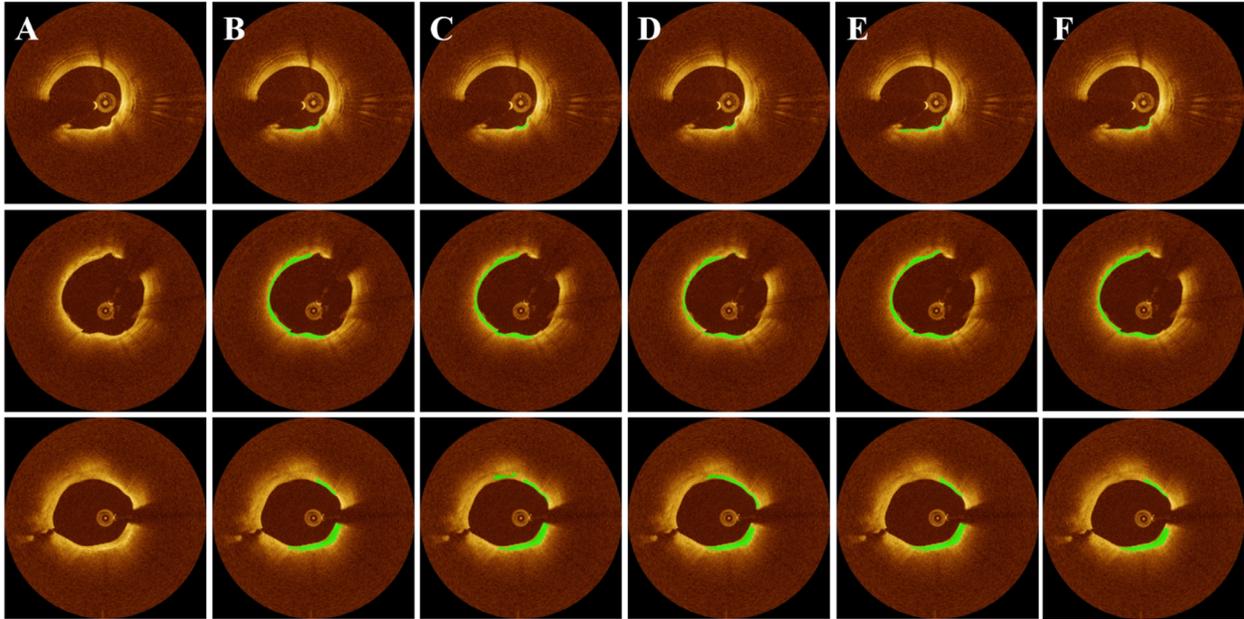

Fig 3. FC Segmentation results for different deep learning models with transfer learning. The panels include (A) IVOCT image in Cartesian coordinates, (B) ground truth, (C) U-Net, (D) Attention U-Net, (E) nnU-Net, and (F) SegResNet. Each row represents different instances of IVOCT images with FC present (shown in green). Among all the networks, SegResNet exhibited the highest segmentation performance in terms of Dice (0.837±0.012) and PPV (82.5%±3.7%) across all five-folds of cross-validation. The green color indicates FC plaque regions.

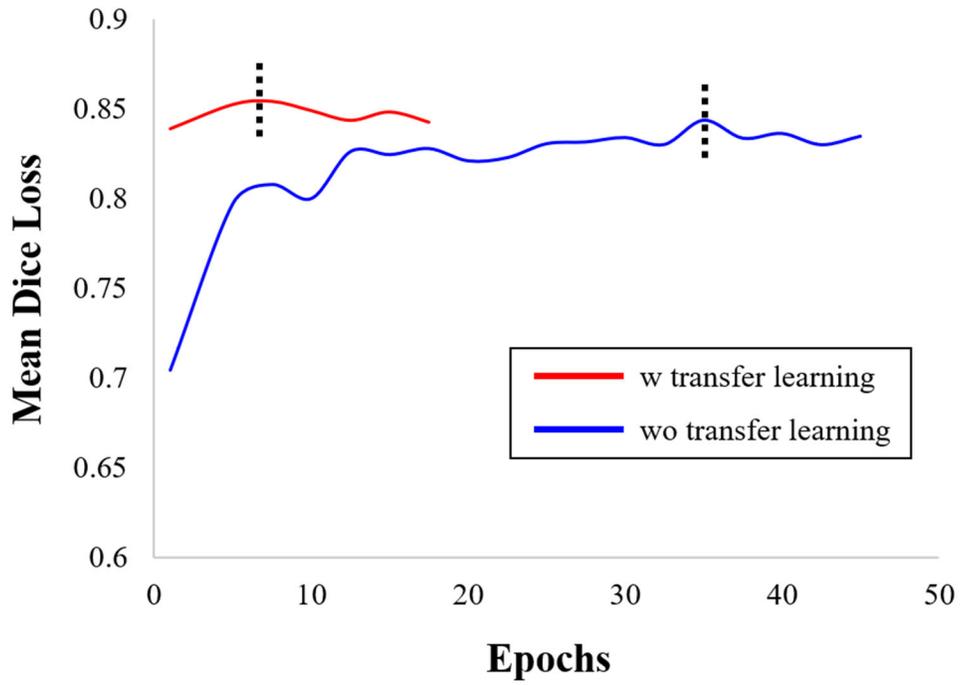

Fig 4. Mean Dice loss curve during validation as computed over a fold. Red and blue curves are results with and without transfer learning, as described in the text. With transfer learning, the curve reached an asymptotic result with many fewer epochs. In addition, there was an improved Dice value in this run with transfer learning. The black dotted lines indicate the points of highest Dice coefficients for each curve.

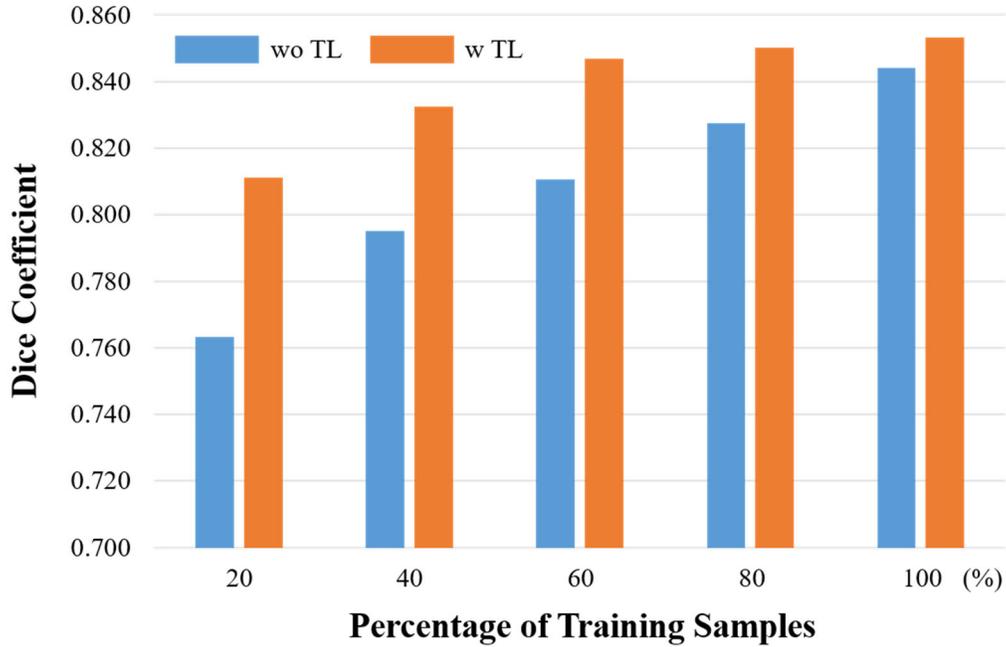

Fig 5. Effect of pretraining and transfer learning on the number of labeled samples required for training. With transfer learning (TL, brown bars), Dice values are always greater than the result without transfer learning. In addition, with transfer learning, there is convergence towards an asymptotic Dice value whereas without transfer learning, performance is continuing to improve much between 80% and 100% of labeled training samples. Note that with transfer learning, only 60% of samples gives a better result than 100% of the samples without transfer learning. The task for pretraining was segmentation of calcified plaques in IVOCT images.

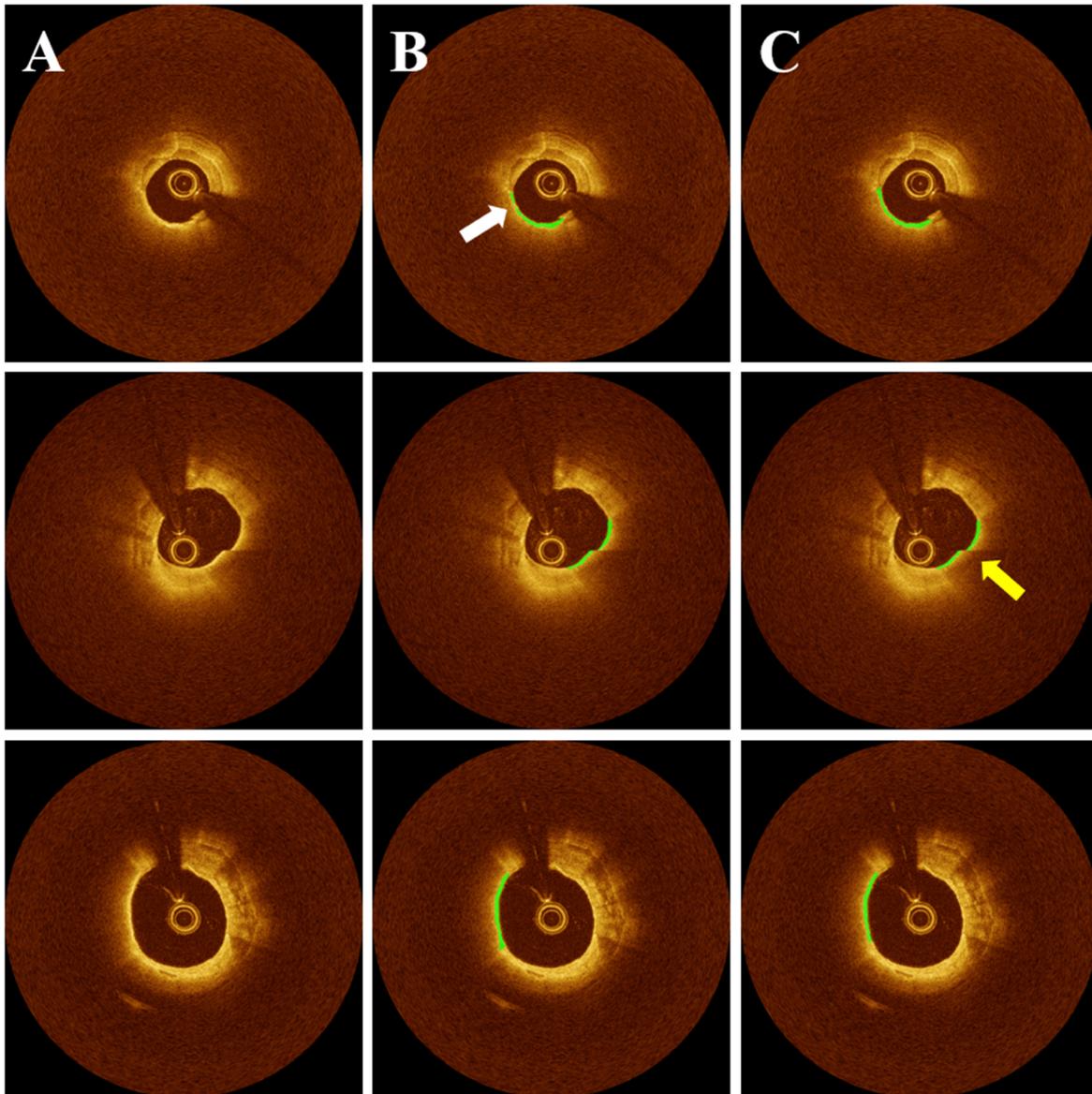

Fig 6. FC segmentation results on the held-out test set. The panels include (A) Cartesian IVOCT image, (B) ground truth, and (C) automated prediction. Each row represents different instances of IVOCT images. In panel B (top), the ground truth FC label appears disconnected at 8 o'clock (indicated by the white arrow); however, our proposed method accurately predicts FC regions, demonstrating its high generalizability. Moreover, our method delivered reliable results even in the presence of image reconstruction errors, as depicted in panel C (at 3 o'clock, highlighted by the yellow arrow). The green color indicates FC plaque regions.

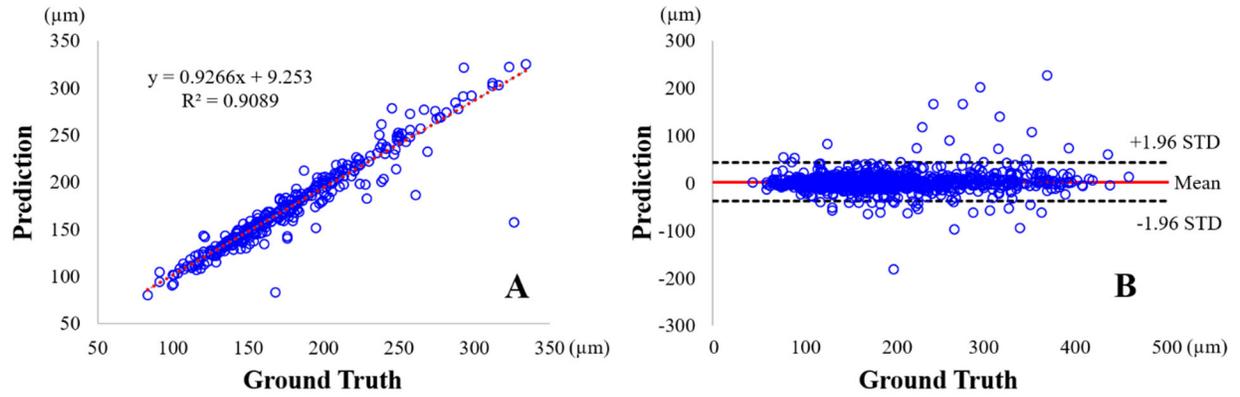

Fig 7. Comparison of mean FC thickness measurements between the ground truth and the proposed method, calculated on the held-out test set. The held-out test set comprised 1,362 FC images from 74 IVOCT pullbacks (UHCMC). In panel (A), the linear regression analysis demonstrates a remarkably high similarity ($R^2$: 0.909) between the proposed method and the ground truth. Additionally, in the Bland-Altman analysis (panel B), the mean bias of FC thickness measurements was 2.95±20.73 $\mu m$, with only a small number of cases (32 out of 1,362) exceeding the limits of agreement (indicated by black dotted lines). These findings indicate the absence of significant bias in the proposed method compared to the ground truth.

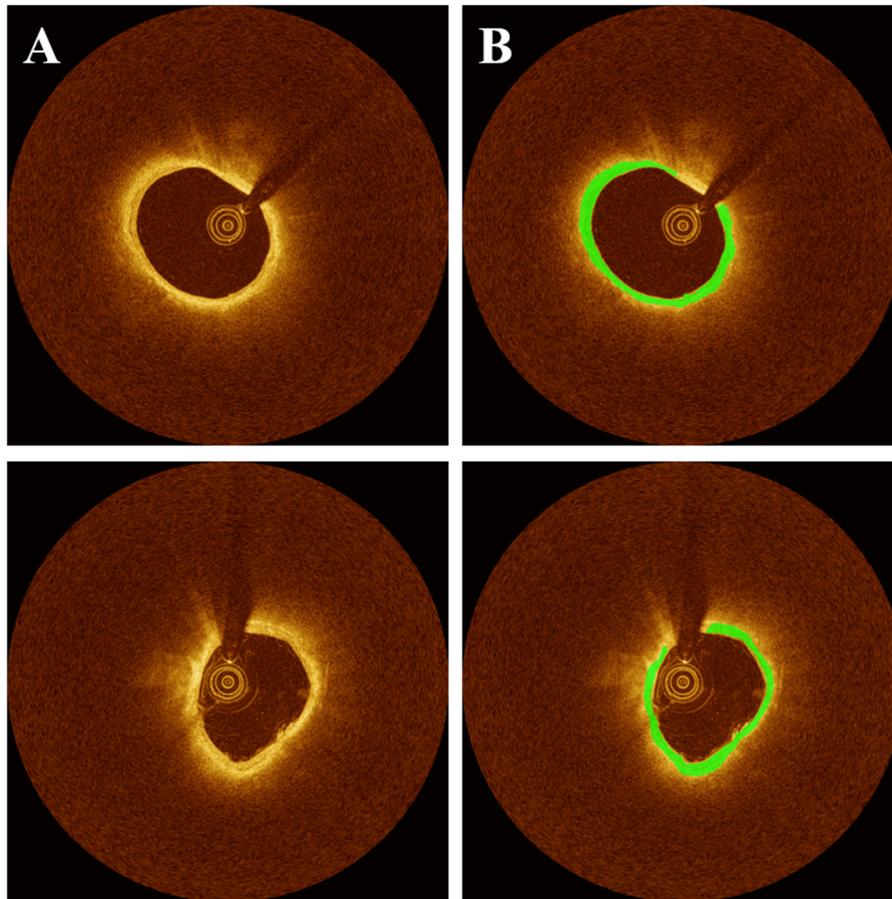

Fig 8. Reproducibility of FC assessment in scan-rescan IVOCT images as obtained from an untreated lesion in paired pre- and post-stenting IVOCT pullbacks. The panels include (A) Cartesian IVOCT image and (B) automated prediction. The top and bottom rows correspond to the first (pre-stenting) and second (post-stenting) scans, respectively. In this case, FC measurements between scans were as follows: FC thickness (114 $\mu m$ / 130 $\mu m$), FC arc angle (314° / 321°), and FC area (1.76 $mm^2$ / 1.49 $mm^2$), indicating its excellent reproducibility. The color green represents FC plaque regions.

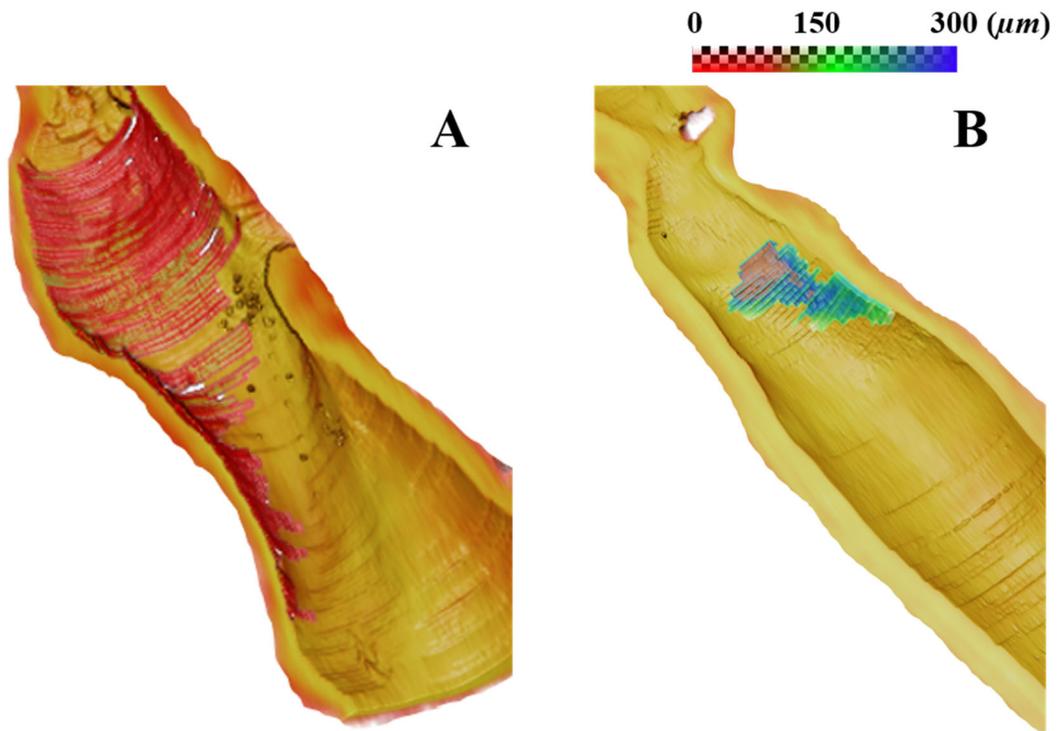

Fig 9. 3D visualizations of FC thickness from representative IVOCT pullbacks with (A) large and (B) small lesions. Both lesions are by definition TCFAs because they have at least a point where the FC is under 65 $\mu m$. In the case of the large lesion, the FC had length = 28.2 $mm$, maximum angle = 271°, and surface area = 66.0 $mm^2$. These values suggest vulnerability as compared to the small lesion with attributes of an FC length of 4.9 $mm$, a maximum angle of 109°, and a surface area of 4.2 $mm^2$.

# Tables

Table 1 Mean quantitative performance metrics of FC segmentation across five-folds for various deep learning networks, including U-Net, Attention U-Net, nnU-Net, and SegResNet. Among all the networks employed, SegResNet demonstrated the best segmentation results, achieving a Dice coefficient of 0.846±0.011 and a PPV of 84.2%±1.8%.

| | PPV (%) | NPV (%) | Sensitivity (%) | Specificity (%) | Accuracy (%) | Dice |
|---|---|---|---|---|---|---|
| U-Net | 84.0±1.8 | 98.9±0.1 | 80.2±0.1 | 99.1±0.1 | 98.1±0.0 | 0.820±0.009 |
| Attention U-Net | 81.2±0.6 | 98.9±0.2 | 80.1±4.9 | 98.9±0.2 | 97.9±0.1 | 0.806±0.022 |
| nnU-Net | 77.2±0.3 | 99.5±0.1 | 91.4±1.5 | 98.5±0.1 | 98.1±0.0 | 0.837±0.008 |
| SegResNet | 84.2±1.8 | 99.1±0.0 | 85.0±0.3 | 99.1±0.1 | 98.3±0.1 | 0.846±0.011 |

Table 2 Quantitative performance metrics of FC segmentation on the held-out test set, including PPV, NPV, sensitivity, specificity, accuracy, and Dice coefficient. Our method demonstrates highly consistent segmentation results between the cross-validation and held-out test sets, highlighting the remarkable generalizability of the proposed approach.

|  | PPV (%) | NPV (%) | Sensitivity (%) | Specificity (%) | Accuracy (%) | Dice |
|---|---|---|---|---|---|---|
| Our Method | 78.5 | 99.4 | 84.9 | 99.0 | 98.5 | 0.816 |

## Supplementary Materials

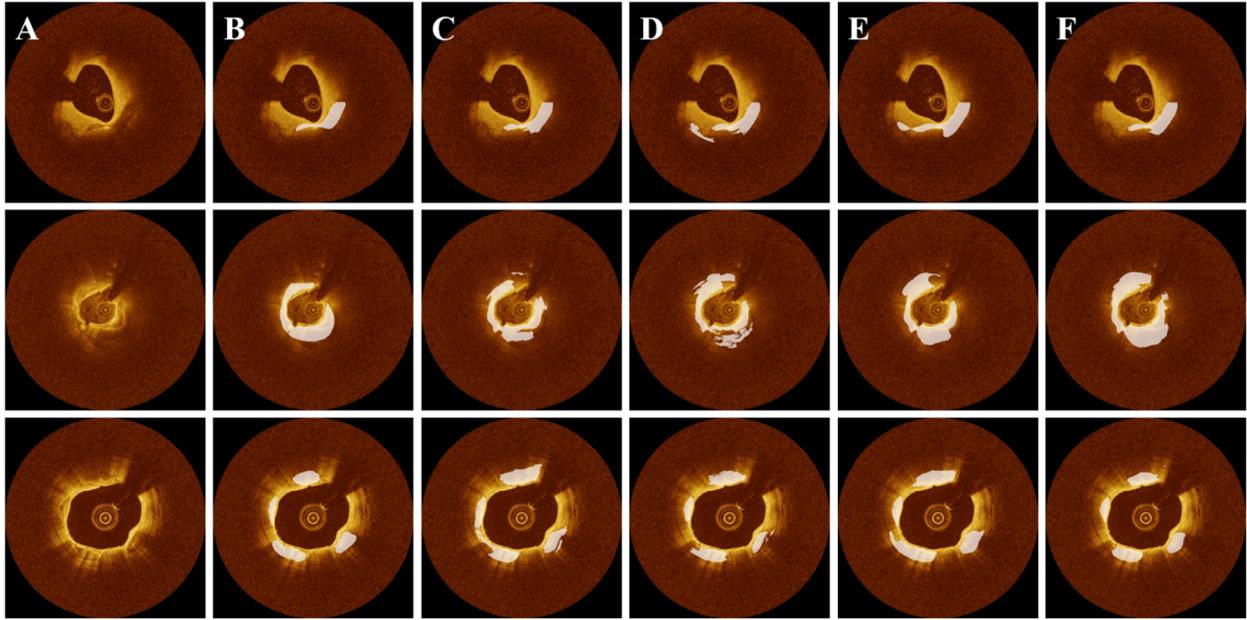

Fig. S1. Automated calcium segmentation results obtained using four different deep learning models. Panels include (A) original image, (B) ground truth, (C) U-Net, (D) Attention U-Net, (E) nnU-Net, and (F) SegResNet. Each row represents different instances of IVOCT images. Overall, SegResNet exhibited the most favorable segmentation results among all the networks employed. The color white indicates areas of calcification.

Table S1 Quantitative performance metrics of calcification segmentation, including PPV, NPV, sensitivity, specificity, accuracy, and Dice coefficient. The networks were trained using existing labeled calcium IVOCT dataset. Among all the networks, SegResNet demonstrated the highest values for PPV, sensitivity, specificity, accuracy, and Dice coefficient, indicating superior performance in calcium segmentation.

|  | PPV (%) | NPV (%) | Sensitivity (%) | Specificity (%) | Accuracy (%) | Dice |
|---|---|---|---|---|---|---|
| U-Net | 77.2 | 97.4 | 76.5 | 97.5 | 95.4 | 0.768 |
| Attention U-Net | 76.3 | 97.2 | 74.6 | 97.4 | 95.2 | 0.755 |
| nnU-Net | 67.4 | 98.6 | 87.4 | 95.3 | 94.5 | 0.761 |
| SegResNet | 79.2 | 97.8 | 80.6 | 97.7 | 96.0 | 0.799 |

The pretrained models for transfer learning yielded reasonable segmentation results for calcification. In the Supplementary Figure 1, the automated calcification predictions from the four different networks on the validation set are depicted. Among them, the Attention U-Net exhibited the largest segmentation errors across most instances, irrespective of the calcification phenotype, and achieved the lowest sensitivity (74.6%) and Dice coefficient (0.755). The U-Net produced similar results. The nnU-Net tended to slightly overestimate coronary calcification, resulting in the lowest PPV (67.4%) despite its high sensitivity (87.4%). On the other hand, the SegResNet demonstrated the most reliable segmentation outcomes, with the highest PPV and Dice coefficient. The Supplementary Table 1 presents the quantitative metrics for calcification segmentation across the four different deep learning networks.

Table S2 Mean quantitative performance metrics of FC segmentation over five folds, comparing results obtained with and without transfer learning (TL). As described in the Results, the application of transfer learning did not yield a significant improvement in segmentation performance. However, it notably reduced the training time for all the networks employed.

|  | PPV (%) | NPV (%) | Sensitivity (%) | Specificity (%) | Accuracy (%) | Dice |
|---|---|---|---|---|---|---|
| U-Net wo TL | 84.3±1.7 | 98.8±0.1 | 79.6±0.3 | 99.1±0.1 | 98.1±0.0 | 0.819±0.006 |
| U-Net w TL | 84.0±1.8 | 98.9±0.1 | 80.2±0.1 | 99.1±0.1 | 98.1±0.0 | 0.820±0.009 |
| Attention U-Net wo TL | 79.2±1.0 | 98.7±0.2 | 77.3±1.5 | 98.8±0.0 | 97.7±0.1 | 0.782±0.003 |
| Attention U-Net w TL | 81.2±0.6 | 98.9±0.2 | 80.1±4.9 | 98.9±0.2 | 97.9±0.1 | 0.806±0.022 |
| nnU-Net wo TL | 80.7±1.5 | 99.3±0.0 | 87.2±0.3 | 98.8±0.1 | 98.2±0.0 | 0.838±0.009 |
| nnU-Net w TL | 77.2±0.3 | 99.5±0.1 | 91.4±1.5 | 98.5±0.1 | 98.1±0.0 | 0.837±0.008 |
| SegResNet wo TL | 84.9±1.8 | 99.0±0.1 | 82.8±0.3 | 99.2±0.1 | 98.3±0.0 | 0.838±0.008 |
| SegResNet w TL | 84.2±1.8 | 99.1±0.0 | 85.0±0.3 | 99.1±0.1 | 98.3±0.1 | 0.846±0.011 |

Table S3 Lesion attributes between pre- and post-stenting pullbacks, including average FC thickness, average FC arc angle, average FC area, and FC surface area. As described in the Results, our method demonstrated excellent reproducibility in automated FC segmentation. STDV represents the standard deviation, and COV represents the coefficient of variance.

|  | Average FC thickness ($\mu m$) | | Average FC arc angle (°) | | Average FC area ($mm^2$) | | FC surface area ($mm^2$) | |
|---|---|---|---|---|---|---|---|---|
|  | Pre | Post | Pre | Post | Pre | Post | Pre | Post |
| Mean | 87.6 | 105.8 | 200.9 | 202.0 | 1.04 | 1.01 | 8.6 | 7.5 |
| STDV | 38.6 | 33.9 | 128.0 | 121.1 | 0.62 | 0.56 | - | - |
| COV | 0.44 | 0.32 | 0.64 | 0.60 | 0.59 | 0.56 | - | - |

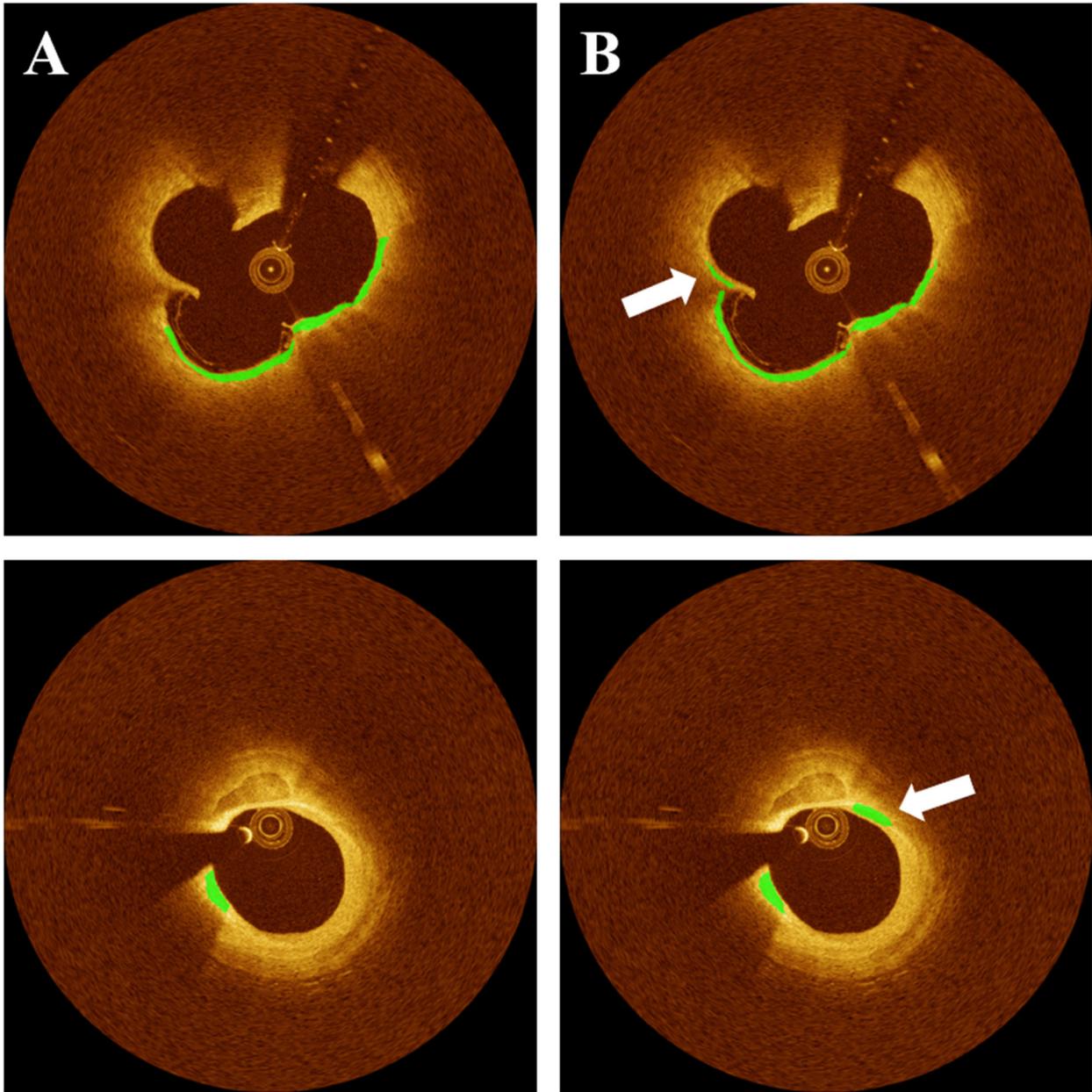

Fig. S2. Illustration of false prediction instances. The panels are (A) manual annotation and (B) automated prediction. In the top example, the proposed method exhibited a false prediction, possibly attributed to the presence of a significant side branch. In the bottom example, the proposed method demonstrated a false prediction concerning a mixed plaque. The false predictions are highlighted with white arrows.